\documentclass[a4paper,12pt,english]{article}
\usepackage[dvipdfm]{color,graphicx}
\DeclareMathAlphabet{\mib}{OML}{cmm} {b}{it}
\definecolor{cyan}{cmyk}{1,0,0,0}
\definecolor{lightcyan}{cmyk}{0.5,0,0,0}
\definecolor{pastelcyan}{cmyk}{0.25,0,0,0}
\definecolor{magenta}{cmyk}{0,1,0,0}
\definecolor{yellow}{cmyk}{0,0,1,0}
\definecolor{lightyellow}{cmyk}{0,0,0.5,0}
\definecolor{pastelyellow}{cmyk}{0,0,0.25,0}
\definecolor{black}{cmyk}{0,0,0,1}
\definecolor{darkgray}{cmyk}{0,0,0,0.75}
\definecolor{gray}{cmyk}{0,0,0,0.5}
\definecolor{lightgray}{cmyk}{0,0,0,0.25}
\definecolor{white}{cmyk}{0,0,0,0}
\definecolor{red}{cmyk}{0,1,1,0}
\definecolor{orange}{cmyk}{0,0.5,1,0}
\definecolor{scarlet}{cmyk}{0,1,0.5,0}
\definecolor{brown}{cmyk}{0.5,0.75,1,0}
\definecolor{camel}{cmyk}{0.25,0.375,0.5,0}
\definecolor{cream}{cmyk}{0,0.2,0.3,0}
\definecolor{green}{cmyk}{1,0,1,0}
\definecolor{lightgreen}{cmyk}{0.5,0,0.5,0}
\definecolor{pastelgreen}{cmyk}{0.25,0,0.25,0}
\definecolor{mossgreen}{cmyk}{0.64,0.4,1,0}
\definecolor{yellowgreen}{cmyk}{0.5,0,1,0}
\definecolor{skyblue}{cmyk}{0.4,0.16,0,0}
\definecolor{royal}{cmyk}{1.0,0.5,0,0}
\definecolor{navyblue}{cmyk}{0.9,0.75,0.5,0}
\definecolor{blue}{cmyk}{1,1,0,0}
\definecolor{lightblue}{cmyk}{0.5,0.5,0,0}
\definecolor{lavender}{cmyk}{0.25,0.25,0,0}
\definecolor{violet}{cmyk}{0.75,1,0.25,0}
\definecolor{purple}{cmyk}{0.5,1,0.5,0}
\definecolor{pink}{cmyk}{0,0.5,0,0}
\definecolor{pastelpink}{cmyk}{0,0.25,0,0}
\def\e{{\rm e}}
\def\gtsim{\mathrel{\hbox{\raise0.2ex
\hbox{$>$}\kern-0.75em\raise-0.9ex\hbox{$\sim$}}}}
\def\ltsim{\mathrel{\hbox{\raise0.2ex
\hbox{$<$}\kern-0.75em\raise-0.9ex\hbox{$\sim$}}}}
\def\llt{\mathrel{<\kern-0.5em <}}   
\def\ggt{\mathrel{>\kern-0.5em >}}   
\def\half{{1\over2}}
\def\kslash{k\kern-0.55em\raise 0.14ex\hbox{/}}
\def\Aslash{A\kern-0.6em\raise 0.14ex\hbox{/}}
\def\dslash{\del\kern-0.55em\raise 0.14ex\hbox{/}}
\def\therefore{ \hskip 0.0pt \raise0.1ex\hbox{.} \mkern -0.24mu \raise 1.1ex\hbox{.} \mkern -0.25mu \raise 0.1ex\hbox{.} \   } 
\def\because{ \hskip 0.0pt \raise 1.1ex\hbox{.} \mkern -0.24mu \raise 0.1ex\hbox{.} \mkern -0.25mu \raise 1.1ex\hbox{.} \  } 

\usepackage{babel}

\newcommand {\beq}{\begin{equation}}
\newcommand {\eeq}{\end{equation}}
\newcommand {\bea}{\begin{eqnarray}}
\newcommand {\eea}{\end{eqnarray}}
\newcommand {\nn}{\nonumber \\}

\newcommand {\m}{\mu}

\newcommand {\pl}{\partial}

\newcommand {\al}{\alpha}
\newcommand {\be}{\beta}

\newcommand {\x}{\xi}

\newcommand {\La}{\Lambda}

\newcommand {\sh}{\theta}   

\newcommand {\om}{\omega}
\newcommand {\Om}{\Omega}
\newcommand {\tauo}{\tau_1}

\newcommand {\del}  {\delta}

\newcommand {\Dcal}{{\cal D}}

\newcommand {\Atil}{{\tilde A}}

\newcommand {\ttil} {{\tilde t}}

\newcommand {\Abar}  {{\bar A}}

\newcommand {\rdot}{\dot{r}}
\newcommand {\rddot}{\ddot{r}}

\newcommand {\xdot}{\dot{x}}

\newcommand {\Xdot}{\dot{X}}

\newcommand {\xddot}{\ddot{x}}

\newcommand {\thdot}{\dot{\theta}}
\newcommand {\thddot}{\ddot{\theta}}

\newcommand {\X}{{\bf X}}
\newcommand {\bx}{{\bf x}}

\newcommand {\ra} {\rightarrow}

\newcommand {\pr}   {{\quad .}}
\newcommand {\com}  {{\quad ,}}
\newcommand {\q}    {\quad}

\newcommand {\nl}    {\newline}
\newcommand {\no}   {\mbox{}}

\newcommand {\PTP}  {Prog.Theor.Phys.}

\begin{document}

\title{
Renormalization Group Approach to Dissipative System
      }
\author{Shoichi~ Ichinose}

\maketitle
\begin{center}\emph{
School of Food and Nutritional Sciences, 
University of Shizuoka,\\
Yada 52-1, Shizuoka 422-8526, Japan\\
Corresponding author: ichinose@u-shizuoka-ken.ac.jp
}\end{center}

\section{Introduction\label{intro}}
In order to understand the { dynamical mechanism} of 
the friction phenomena, we heavily rely on the numerical 
analysis using various methods: molecular dynamics, Langevin equation, 
lattice Boltzmann method, Monte Carlo, e.t.c.. 
For the case of the Langevin equation, for example,  
a simple model is described as follows.    
\bea
m\xddot=-{ \eta}\xdot-\frac{\pl V}{\pl x}+{ \mbox{random force}}\com\nn
{ \eta}\ :\ \mbox{viscosity},\ \ 
V(x):\ \mbox{potential}  
\pr
\label{int1}
\eea 
The effect of the random force is given by the white noise. 
We propose a new method which has the following characteristic points:\ 
1)\ the {\it geometrical} approach to the statistical mechanical system;\ 
2)\ the {\it continuum} approach using Feynman's path integral (generalized version);\ 
3)\ the {\it holographic} (higher-dimensional) approach;\  
4)\ the {\it renormalization} phenomenon takes place in order to treat the statistical fluctuation.

In ref.\cite{ICSF2010, SI1004}, we have explained this method using the above model (\ref{int1}).

\section{One Dimensional Spring-Block Model\label{block}}
\begin{figure}
\caption{
One dimensional spring-block model
        }
\begin{center}
\includegraphics[height=6cm]{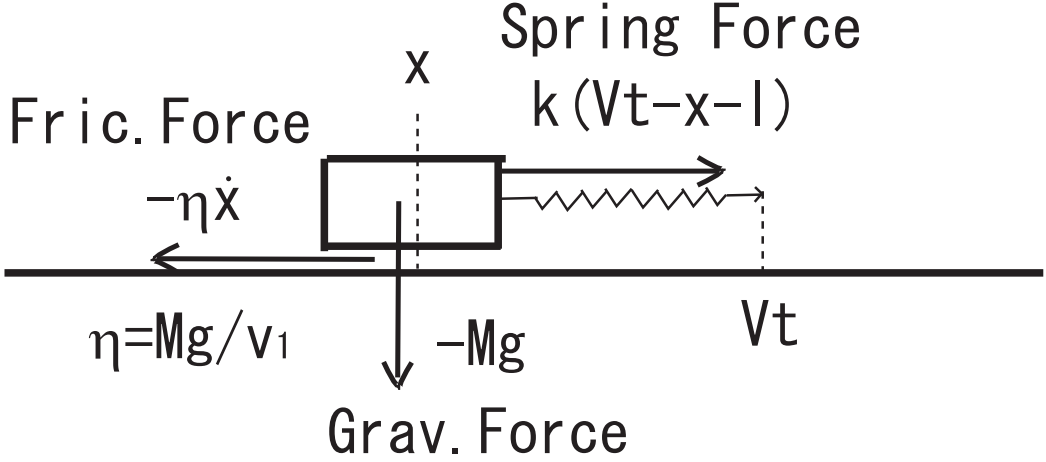}
\end{center}
\label{SBmodel}
\end{figure}

We take another simple model of the friction system: the spring-block model. 
See Fig.\ref{SBmodel}. 
It describes the movement of a block (rigid body), dragged by the spring, 
on a table with friction. The front end of the spring moves with the constant velocity 
$V$. 
The equation of motion is given by 
\bea
M\xddot=-\mu(\xdot)Mg+k(Vt-x-l),\ 
\mu(\xdot)=\frac{1}{v_1}\xdot
\ ,
\label{block1}
\eea 
where 
$M$ is the {mass},\ $\mu$ is the {friction coefficient},\  
$g$ is the {gravitational acceleration constant},\  
$k$ is the {spring constant},\ $V$ is the {front-end velocity (constant)},\ and   
$l$ is the {natural length of the spring}. 
It can be re-written as 
\bea
\xddot+\frac{1}{\tau_1}\xdot +\om^2 x=\om^2 (Vt-l),\ \tau_1\equiv\frac{v_1}{g},\ 
\om\equiv \sqrt{\frac{k}{M}}
\ .
\label{block3}
\eea 

\begin{figure}
\caption{
The solution (\ref{block4}) with $\tau_1=V=l=1.0, \om=10.0, \Om=\sqrt{399}/2, 0\leq t \leq 2$. 
        }
\begin{center}
\includegraphics[height=8cm]{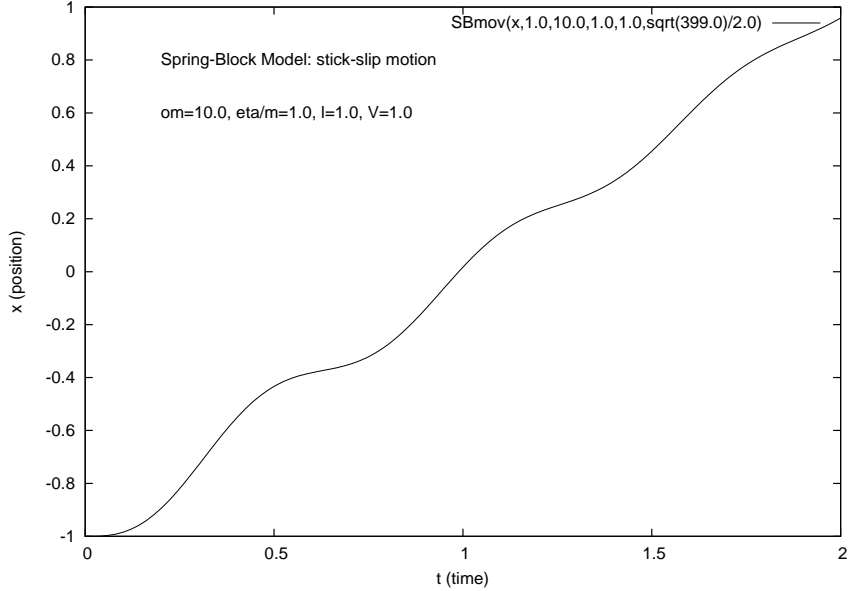}
\end{center}
\label{StickSlip}
\end{figure}

For the initial condition:\ $x(0)=-l\ ,\ \xdot(0)=0$, the classical solution 
(elasticity dominate case, $4\om^2 > 1/{\tauo}^2$) is given by 
\bea
x(t)=\e^{-t/2\tauo} V \{ (1/2\om^2\tauo^2-1)(\sin\Om t)/\Om +(1/\om^2\tauo)\cos\Om t \}\nn
-l+V (t-1/\om^2\tauo),\ \ 
\Om=(1/2)\sqrt{4\om^2-1/{\tauo}^2},\nn 
0\leq t\leq 2,\ x(0)=-l,\ \xdot(0)=0
.
\label{block4}
\eea 
See Fig.\ref{StickSlip}. It shows the 'stick-slip' motion. The time interval 
for one pair of the stick and slip-state is $2\pi/\Om$. 

We impose the {\it periodicity} on the t-axis 
for the (IR) regularization. (Physically this procedure is regarded as putting the whole system
in the heat-bath of the temperature $\be^{-1}$. )
\bea
t\ \ra\ t+\be,\ \be^{-1}:\ \mbox{temperature}
.
\label{block4b}
\eea 

The equation of motion (\ref{block3}) gives us the following relation. 
\bea
\left[
\half\xdot^2+\frac{\om^2}{2}x^2+\om^2 lx
\right]_{t_1}^{t_2}=-\frac{1}{\tau_1}\int_{t_1}^{t_2}\xdot^2 dt+ \om^2 V \int_{t_1}^{t_2}t\xdot dt
.
\label{block5}
\eea 
Changing variables $t_2$ to $t$, and $t_1$ to $0$, we read 
the {\it energy conservation equation}. 
\bea
H[\xdot,x]\equiv\half\xdot^2+\frac{\om^2}{2}x^2+\om^2 lx
+\frac{1}{\tau_1}\int_{0}^{t}\xdot^2 d\ttil- \om^2 V \int_{0}^{t}\ttil\xdot d\ttil\nn
=\left( \half\xdot^2+\frac{\om^2}{2}x^2+\om^2 lx \right)_{t=0}=-\frac{\om^2}{2}l^2
=E_0\ (\mbox{constant})
\com
\label{block6}
\eea 
where we have used the initial condition: $x(0)=-l\ ,\ \xdot(0)=0$. 
In the second formula, the fourth and the fifth terms are the {\it hysteresis} ones, 
$\{ x(\ttil):\ 0\leq \ttil \leq t \}$. 
The fourth term is the {\it friction-heat} energy produced until the time $t$. 
The fifth one 
is the {\it subtraction} of the cumulated external work done by the dragging until the time $t$.   
From this Hamiltonian (energy) expression, (\ref{block6}), we can read the (bulk) {\it metric} 
in the 2 dimensional (D) space $(X,t)$.\cite{ICSF2010} There are two types. \nl 
 
\vspace{3mm}
{\large {Dirac Type}}\nl    
\vspace{3mm}
\bea
ds^2\no_D\equiv
dX\no^2+(\om^2 X^2+2\om^2 lX)~dt\no^2\nn
+dt\no^2\left\{
\frac{2}{\tau_1}\int_{-l}^{x(t)}\Xdot dX- 2\om^2 V \int_{-l}^{x(t)}\ttil~ dX
\right\}
,
\label{block7}
\eea 
where $0\leq \ttil \leq t$ and $\Xdot = \frac{dX}{d\ttil}$. 
From this construction of the bulk metric, we see $ds^2\no_D$ 
reduces, on a path $X=x(t)$, to be proportional to the {\it energy}. 
On a path
\bea
X=x(t),\ 0\leq t \leq \be,\ dX=\xdot dt
,
\label{block8}
\eea 
the {\it induced} metric is given by 
\bea
ds^2\no_D|_{on-path}=\left[
\xdot^2+\om^2 x^2+2\om^2 lx   \right.\nn
\left. +\frac{2}{\tau_1}\int_{0}^{t}\xdot^2 d\ttil- 2\om^2 V \int_{0}^{t}\ttil\xdot d\ttil
                              \right]dt\no^2
\equiv 2H[\xdot,x] dt\no^2
.
\label{block9}
\eea 
The {\it length} $L$ of the path $\{ x(t) | 0\leq t \leq \be\}$ is given by 
\bea
L[x(t)]=\int ds=\int_0^\be\sqrt{2H}dt
.
\label{block10}
\eea 
\nl
\vspace{3mm}
{\large {Standard Type}}\nl    
\vspace{3mm}
We take the following form for the line element $ds^2\no_S$.  
\bea
ds^2\no_S=\frac{1}{dt\no^2}(ds^2\no_D)^2\ -\ \mbox{on-path}\ \ra\ 
(2H[\xdot,x])^2 dt\no^2
.\label{block11}
\eea 
On the path, the length is given by
\bea
L[x(t)]=2\int_0^\be H[\xdot, x] dt
.
\label{block12}
\eea 

\begin{figure}
\caption{
The paths appearing in the path-integral expression (\ref{block13}) 
of the free energy during the movement of the block. 
        }
\begin{center}
\includegraphics[height=6cm]{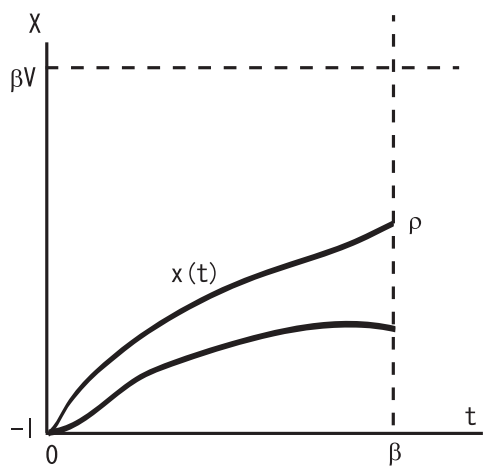}
\end{center}
\label{SBpath}
\end{figure}

The spring-block starts with the stick-slip motion and finally 
reaches the {\it steady state}  with a constant velocity. 
During the movement the friction-heat is produced and 
the external work, by the dragging, keeps being given. 
Microscopically the statistical fluctuation 
occurs in the (bottom) surface of the block. 
For both types of metric, 
the free energy $F$ is given by the path-integral \cite{Fey72} with 
the {\it statistical weight} $\exp (-L/2\al')$
(the statistical ensemble measure based on the geometry\cite{ICSF2010,SI0801}). 
\bea
\e^{-\be F[l,\al', \be; \om, \tau_1, V]}=\int_{-l}^{\be V}d\rho
\int_{\begin{array}{c}x(0)=-l\\x(\be)=\rho\end{array}}\nn
\times\prod_t \Dcal x(t)\exp \left[-\frac{1}{2\al'}L[x(t)]\right]
.
\label{block13}
\eea 
The paths are shown in Fig.\ref{SBpath}. 
$\al'$ is a {\it new} model parameter which shows the tension of the 1D 
string (line). 
The parameter $\al'$ has the dimension of length ($[\al']=\mbox{L}$) for Dirac-type, 
while that of Standard-type has the dimension of length$\times$velocity 
($[\al']\ ={\mbox{L}}^2/{\mbox{T}} $). $\be^{-1}$ is regarded as the temperature 
of the final (equilibrium) state. 

The energy E, and the entropy S are obtained by
\bea
\mbox{Energy}\ E(l,\al', \be; \om, \tau_1, V)=<\frac{L}{2}>=
\frac{\pl}{\pl (-{\al'}^{-1})}\e^{-\be F},  \nn
\mbox{Entropy}\q S(l,\al', \be; \om, \tau_1, V)=k\be (E-F).
\label{central10}
\eea 
The effective {\it force} emerges, as the statistically averaging effect, 
in the spring-block and is given by 
\bea
\mbox{Force}\q f(l,\al', \be; \om, \tau_1, V)=-\frac{\pl E}{\pl l}
\label{central11}
\eea 

Evaluation of the free energy, (\ref{block13}), requires the {\it renormalization} 
of some parameters. 

\section{Two Dimensional Dissipative Block in the Central Force\label{central}}

\begin{figure}
\caption{
2D dissipative block in the central force
        }
\begin{center}
\includegraphics[height=6cm]{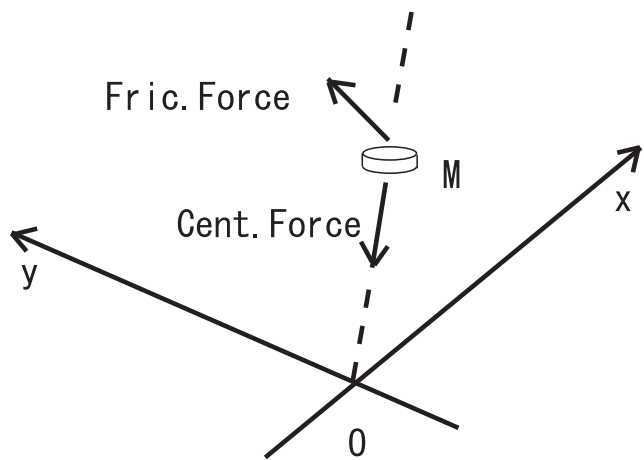}
\end{center}
\label{2DSBmodel}
\end{figure}

We consider the case that a block moves on a plane, 
described by the coordinates $\bx=(x,y)$, with the friction under the influence of 
the central force. 
\bea
M\ddot{\bx} =-\mu(\dot{\bx})Mg\frac{\dot{\bx}}{|\dot{\bx}|}+F(|\bx|)\frac{\bx}{|\bx|},
M:\mbox{mass};\ g:\mbox{grav. accel.}  \nn
\mbox{Fric. Coeff.};\   
\mu(\dot{\bx})=\frac{|\dot{\bx}|}{v_1},\ 
\mbox{Cent. Force}:\ F=-k|\bx|.
\label{central1}
\eea 
The classical equation of motion is expressed as 
\bea
\ddot{\bx} +\frac{1}{\tau_1}\dot{\bx}+\om^2\bx=0,\ 
\om^2=\frac{k}{M},\ \tau_1=\frac{v_1}{g}
.
\label{central1b}
\eea 
In terms of the polar coordinates $(r,\sh)$, this equation is re-written as  
\bea
\bx = (x=r\cos\sh, y=r\sin\sh)\com\nn
\rddot+\frac{1}{\tau_1}\rdot -r (\thdot^2-\om^2)=0\com\q
r\thddot +(2\rdot +\frac{r}{\tau_1})\thdot=0\pr
\label{central1c}
\eea 

\begin{figure}
\caption{Movement of 
2D spring-block model. $\tau_1=1, \om=1;\ \thdot (0)=\sqrt{3/4}, \sh (0)=0, r(0)=1, 
\rdot (0)=(\sqrt{3}-1)/2.\ \ -0.2\leq x\leq 1.2,\ \ -0.1\leq y\leq 0.6 $
        }
\begin{center}
\includegraphics[height=6cm]{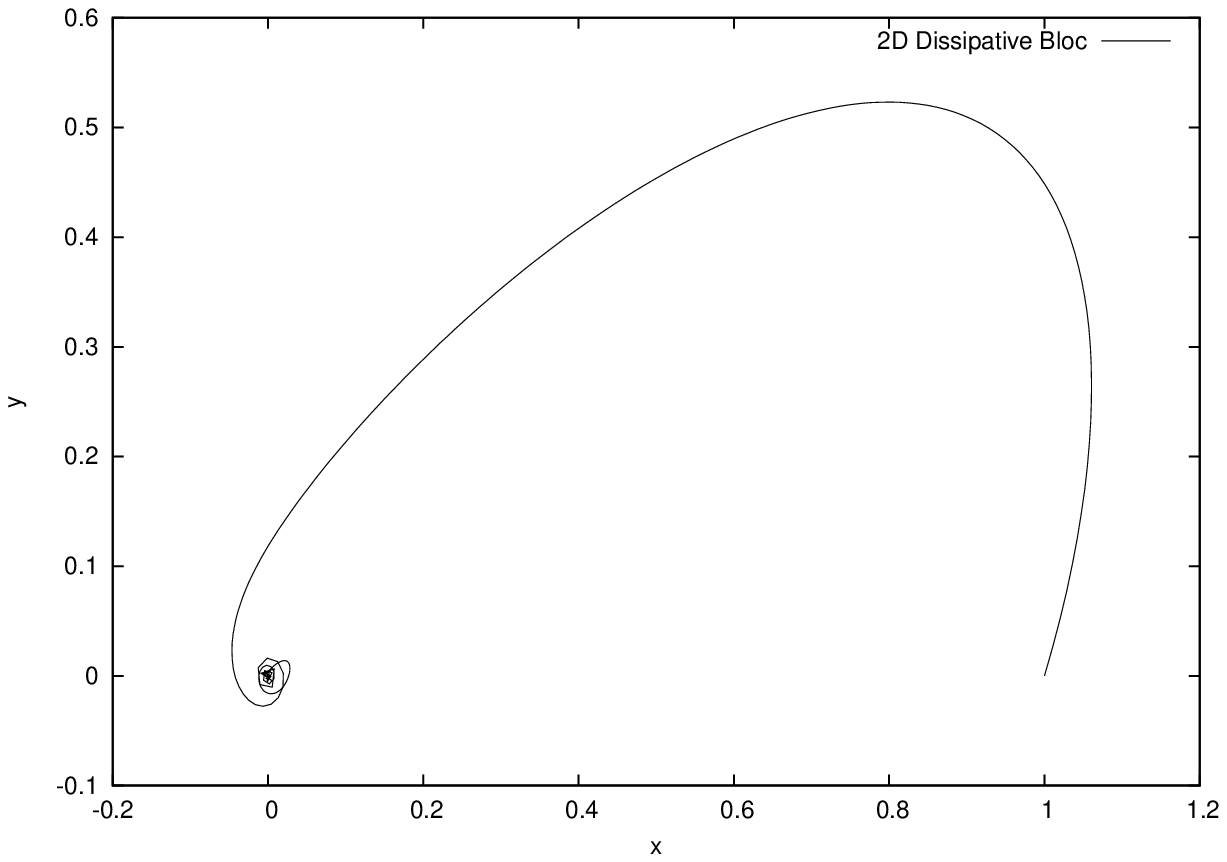}
\end{center}
\label{2dSBmove}
\end{figure}

\begin{figure}
\caption{Movement of 
2D spring-block model. $\tau_1=1, \om=1;\ \thdot (0)=\sqrt{3/4}, \sh (0)=0, r(0)=1, 
\rdot (0)=(\sqrt{3}-1)/2.\ \ -0.1\leq x\leq 0.1,\ \ -0.05\leq y\leq 0.1 $
        }
\begin{center}
\includegraphics[height=6cm]{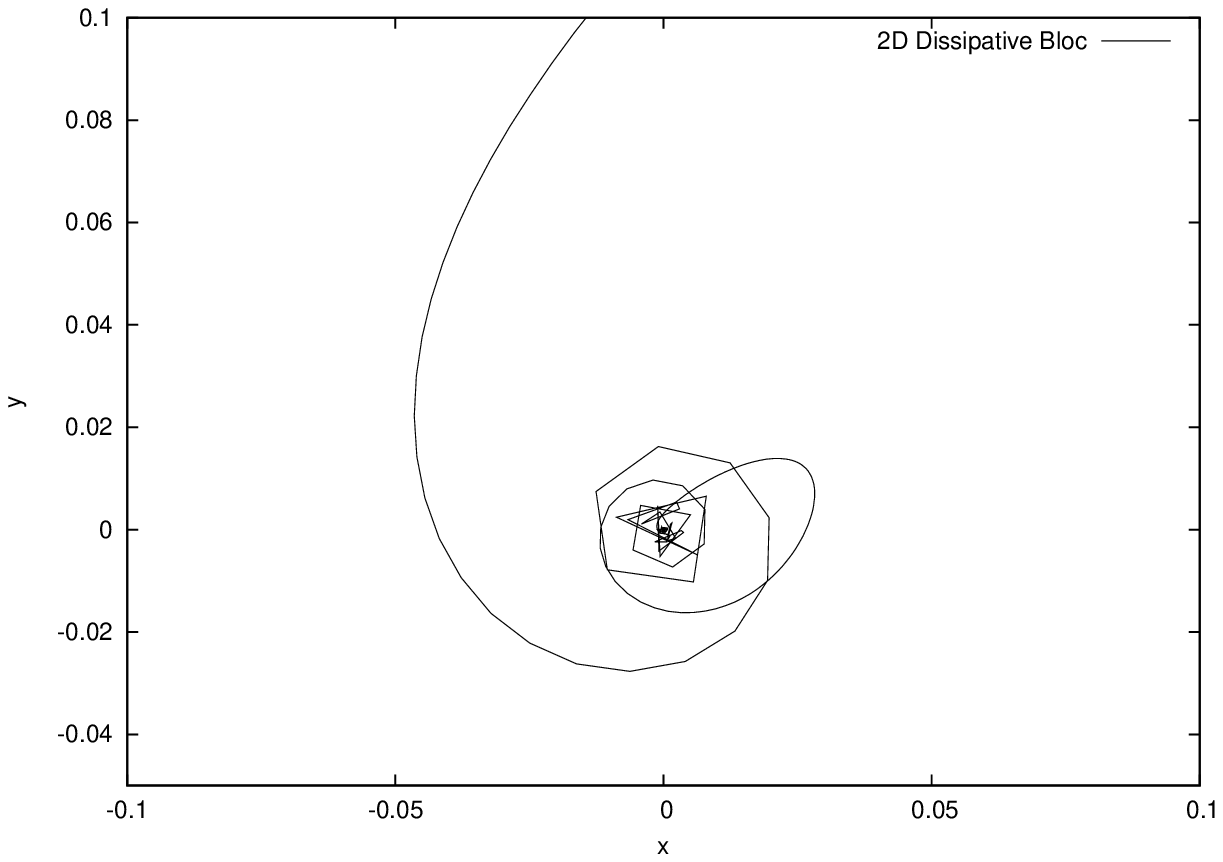}
\end{center}
\label{2dSBmoveNearO}
\end{figure}

A solution (central-force dominant case: $\om^2> \frac{1}{4{\tau_1}^2}$) is given by
\bea
r(t)=\e^{-\frac{t}{2\tau_1}}r_0 (\sin\om_0t\ + 1)\com\nn
\sh(t)=\half\tan (\frac{\om_0t}{2}-\frac{\pi}{4})+\frac{1}{6}\tan^3(\frac{\om_0t}{2}-\frac{\pi}{4})+\frac{2}{3}
\com
\label{central1d}
\eea 
where $r_0$ and $\om_0$ are the parameters appearing in the initial condition 
$r(0)=r_0, \sh(0)=\sh_0, v_r(0)={v_r}^0$ and $\thdot(0)=\om_0$ which we choose 
in the following way.
\bea
\om_0=\sqrt{\om^2-\frac{1}{4{\tau_1}^2}}\com\q
\frac{{v_r}^0}{r_0}=-\frac{1}{2\tau_1}+\sqrt{\om^2- \frac{1}{4{\tau_1}^2}}\com\q
\sh_0=0
\pr
\label{central1e}
\eea 
The choice (\ref{central1e}) is only for the simple form of (\ref{central1d}). 
For the case $\tau_1=1, \om=1$ and $r_0=1$, the orbit is shown in Fig.\ref{2dSBmove} 
for $-0.2\leq x\leq 1.2$ and in Fig.\ref{2dSBmoveNearO} for $-0.1\leq x\leq 0.1$.   
The orbit shows  'stick-slip' motion and 
the time interval for the one motion is $2\pi/\thdot(0)$. 


As in Sec.\ref{block}, we impose the periodicity on t-axis:\ $t\ra t+\be$. 
From the equation of motion (\ref{central1b}) we can derive the following relation.  
\bea
\left[
\half{\dot{\bx}}^2+\frac{\om^2}{2}\bx^2
\right]^t_{0}=-\frac{1}{\tau_1}\int^t_{0}{\dot{\bx}}^2d\ttil,\ \om^2=\frac{k}{M},\ \tau_1=\frac{v_1}{g}
,
\label{central2}
\eea 
where $0\leq t\leq \be$. 
From this result, we obtain the {\it energy conservation equation}. 
\bea
H[\dot{\bx},\bx]\equiv\half{\dot{\bx}}^2+\frac{\om^2}{2}\bx^2
+\frac{1}{\tau_1}\int^t_{0}{\dot{\bx}}^2d\ttil
=\left( \half{\dot{\bx}}^2+\frac{\om^2}{2}\bx^2 \right)|_{t=0}= E_0,
\label{central3}
\eea 
where the third term of the second formula shows the {\it hysteresis} effect and 
$\dot{\bx}=d\bx(\ttil)/d\ttil\ ,\ 0\leq \ttil\leq t$. 
From the above equation, we can read the 3 dim (bulk) {\it metric}. 
\bea
\mbox{Dirac}:\ 
ds^2\no_D\equiv d\X\no^2+\om^2\X^2dt^2
+\frac{2}{\tau_1}dt^2\int^{\X(t)}_{\X_0}\dot{\X}\cdot d\X ,\nn
\mbox{Standard}:\ 
ds^2\no_S\equiv \frac{1}{dt^2}(ds^2\no_D)^2,
\label{central4}
\eea 
where $\dot{\X}=\frac{d\X (\ttil)}{d\ttil},\ 0\leq\ttil\leq t,\ \X_0\equiv \X(0)$. 
The friction occurs between the top surface of the plane and the bottom surface of the block. 
The solution (\ref{central1d}) shows that the block does  
the {\it stick-slip} motion. The friction is microscopically 
caused by the irregularly-distributed {\it asperity} on both surfaces. 
We introduce the {\it distribution} of the movement-configuration 
in the {\it geometrical} way. 
In order to define it, we first prepare 
the following 2D surface in the 3 dimensional (bulk) space ($t, X, Y$). 
\begin{figure}
\caption{
2D surface embedded in 3D space-time by the closed-string condition (\ref{central5}). 
        }
\begin{center}
\includegraphics[height=6cm]{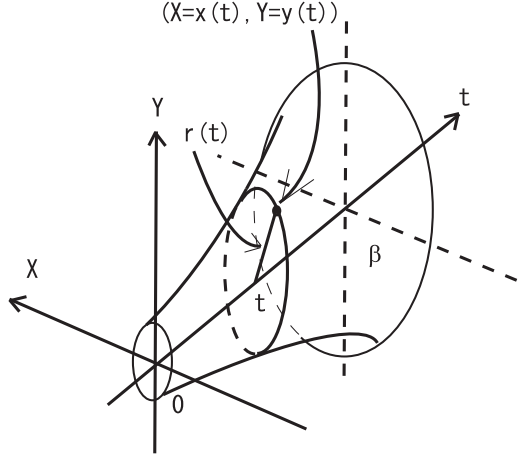}
\end{center}
\label{2DHyperSurf}
\end{figure}
\bea
\mbox{Closed-String Condition}:\nn
\bx^2(t)=x(t)^2+y(t)^2=r(t)^2,\ 
0\leq t\leq \be,
\label{central5}
\eea 
where we assume the 2D world described by the coordinates ($x, y$) is 
{\it isotropic} around the origin. See Fig.\ref{2DHyperSurf}. 
The function form of $r(t)$ can be taken in the arbitrary way. 
The configuration is a {\it closed-string}. $\be$ is a boundary parameter of $t$-axis. 
It plays the role of the inverse {\it temperature} of the final (equilibrium) state of the system. 
The metric on the surface, called "{\it induced metric}", is obtained by 
using the closed-string condition (\ref{central5}) in the expression 
(\ref{central4}). 
\bea
ds^2\no_D|_{on-path}=\nn
dr^2+r^2d\sh^2+\frac{dr^2}{\rdot^2}
\left\{ \om^2r^2+\frac{2}{\tau_1}\int_0^t(\rdot^2+r^2\thdot^2)d\ttil \right\}\nn
\equiv g^{ind}_{ij}dx^idx^j,\q i,j=1,2;\ (x^1,x^2)=(r,\sh)
,
\label{central6}
\eea 
where $x(t)=r(t)\cos\sh(t) ,\ y(t)=r(t)\sin\sh(t)$. 
The {\it induced} metric is explicitly given by
\bea
g^{ind}_{ij}(r(t),\sh(t))=
                                 \left(
\begin{array}{cc}
\frac{2H_r}{\rdot^2}   &  0\\
0  &  \frac{2H_\sh}{\thdot^2}
\end{array}
                                   \right),\q 
2H[r, \rdot, \thdot]=\nn
\rdot^2+r^2\thdot^2+
\om^2r^2+\frac{2}{\tau_1}\int_0^t(\rdot^2+r^2\thdot^2)d\ttil
=2H_r + 2H_\sh \ ,
\label{central7}
\eea 
where $2H_\sh\equiv r^2\thdot^2$ is the angular part of Hamiltonian and 
$2H_r\equiv 2H-2H_\sh$ is the radial one. 
The {\it area} $A$ of the surface is given by
\bea
A[r(t), \sh(t)]=\int\sqrt{\mbox{det}~g^{ind}_{ij}}~d^2x=\nn
\int_0^\be dt \int d\sh(t)\Atil[r(t), \sh(t)],\ 
\Atil=2~\frac{\sqrt{H_r H_\sh}}{|\thdot|}
,
\label{central8}
\eea 
where $d^2x=dr~d\sh=\rdot d\sh dt$. 

The macroscopic physical quantities, such as the energy of the whole system, 
are generally given by the form of the integral over the whole space-time (bulk space). 
They are often divergent\cite{SI0801,SI0812,SI1205}. 
In order to {\it regularize} the singular behavior, and to take the {\it statistical average} at the same time, 
we replace the integral by the sum (integral) over all possible {\it surfaces} satisfying 
the given boundary condition. 
The measure of the path (surface)-integral is taken as follows. 
An infinitesimal surface between $t$ and $t+dt$ is specified by $dr(t)=\rdot dt$ and 
$d\sh$. 
For simplicity, we consider the case: $\thdot=0, \sh=$constant. 
(The angular variable $\sh$ is commonly used for all $t$.  )
The free energy $F$ 
 is given by
\bea
\e^{-\be F[\rho_0, \al', \be; \om, \tau_1]}
=\int_{\La^{-1}}^{\mu^{-1}}d\rho
\int_{\begin{array}{c}r(0)=\rho_0\\r(\be)=\rho\end{array}}
\prod_{t'} r(t')\Dcal r(t') \nn
\times
\exp \left[-\frac{2\pi}{2\al'}\int_0^\be \Abar[r(t)] dt  \right], 
\Abar[r(t)]=r~\sqrt{\rdot^2+\om^2 r^2+\frac{2}{\tau_1}\int_0^t \rdot^2 d\ttil \ }.
\label{central9}
\eea 
where $\La^{-1}$ and $\m^{-1}$ are the UV and IR {\it regularization} parameters. 
$\rho_0$ is a model parameter and shows the starting radial position. 
$\al'$ is the parameter which shows the tension of the embedded 2D surface, Fig.\ref{2DHyperSurf}. 
Its physical dimension is\ \ [$\al'$]=L$^2$.  
The energy E, and the entropy S are obtained by
\bea
\mbox{Energy}:\ E(\rho_0, \al', \be; \om, \tau_1)=<\frac{A}{2}>=
\frac{\pl}{\pl (-{\al'}^{-1})}\e^{-\be F}
,\nn
\mbox{Entropy}:\ S(\rho_0, \al', \be; \om, \tau_1)=k\be (E-F).
\label{central10b}
\eea 
As the statistically averaging effect, 
the effective force emerges, in the radial direction, on the block and is given by 
\bea
\mbox{Force}:\ f(  \rho_0, \al', \be; \om, \tau_1    )=-\frac{\pl E}{\pl \rho_0}\ .
\label{central11b}
\eea 
Evaluation of the free energy F, (\ref{central9}), requires the {\it renormalization} 
of some parameters.

\section{Conclusion\label{conc}}

Recently another new approach to the dissipative system is proposed\cite{SI1303}, 
where the time development is replaced by the step-wise process. The present 
geometric approach is also applied to the condensed matter physics such as 
the permittivity of the substance\cite{SI1203}. 

We have proposed a new formalism to calculate 
the fluctuation effect in the dissipative system based on 
the geometry appearing in the system energy expression. 
The integration measure for the statistical ensemble is 
taken from Feynman's idea of the path-integral. 
It clarifies the statistically-averaging procedure.

\end{document}